\documentclass[aps,prl,twocolumn,superscriptaddress]{revtex4}
\usepackage{graphicx}

\usepackage{dcolumn}

\usepackage{bm}
\usepackage{color}
\usepackage[colorlinks]{hyperref}
\input{epsf}

\begin{document}


\title{Comment on cond-mat/0409228 "Microwave photoresponse in the 2D electron system caused by
intra-Landau level transitions"}

\author{R. G. Mani}
\email{mani@deas.harvard.edu} \affiliation {Harvard University,
Gordon McKay Laboratory of Applied Science, 9 Oxford Street,
Cambridge, MA 02138, USA}
%
%
%
%
\date{\today}
\begin{abstract}
We provide an article-extract which points out that a
microwave-induced modification in the resistance occurs at
relatively "high" magnetic fields where the radiation is incapable
of producing inter-Landau level excitations and, therefore, that
\textit{the microwave radiation must be producing intra Landau
level excitations as well}.
\end{abstract}
%
%
\maketitle In the above mentioned cond-mat preprint entitled
"Microwave photoresponse in the 2D electron system caused by
intra-Landau level transitions", Dorozhkin et al.\cite{1} have
discussed a regime of the microwave-induced response in the 2DES
that does not obtain much attention - the limit of "high" magnetic
fields, $B$, where the Landau level spacing $\hbar \omega_{c}$
exceeds the photon energy $hf$. Based on the available models for
the microwave-induced magnetoresistance,\cite{2} one naively
expects little change in the magnetoresistance upon microwave
excitation in this limit because inter Landau level excitations
due to single photon processes are energetically unfavorable and
unlikely when $B$ $>$ $B_{f}$, i.e., $\hbar\omega_{C}$ $>$ $hf$.
Here, $B_{f}$ = $2\pi f m^{*} /e$, $f$ is the microwave frequency,
$m^{*}$ is an effective mass, and $e$ is the electron charge.
Thus, this $B$ $>$ $B_{f}$ regime is noteworthy because it points
out further complexity in the experimentally observed phenomena,
than what has been appreciated thus far.
\begin{figure}
\begin{center}
\includegraphics*[scale = 0.25,angle=0,keepaspectratio=true,width=7in]{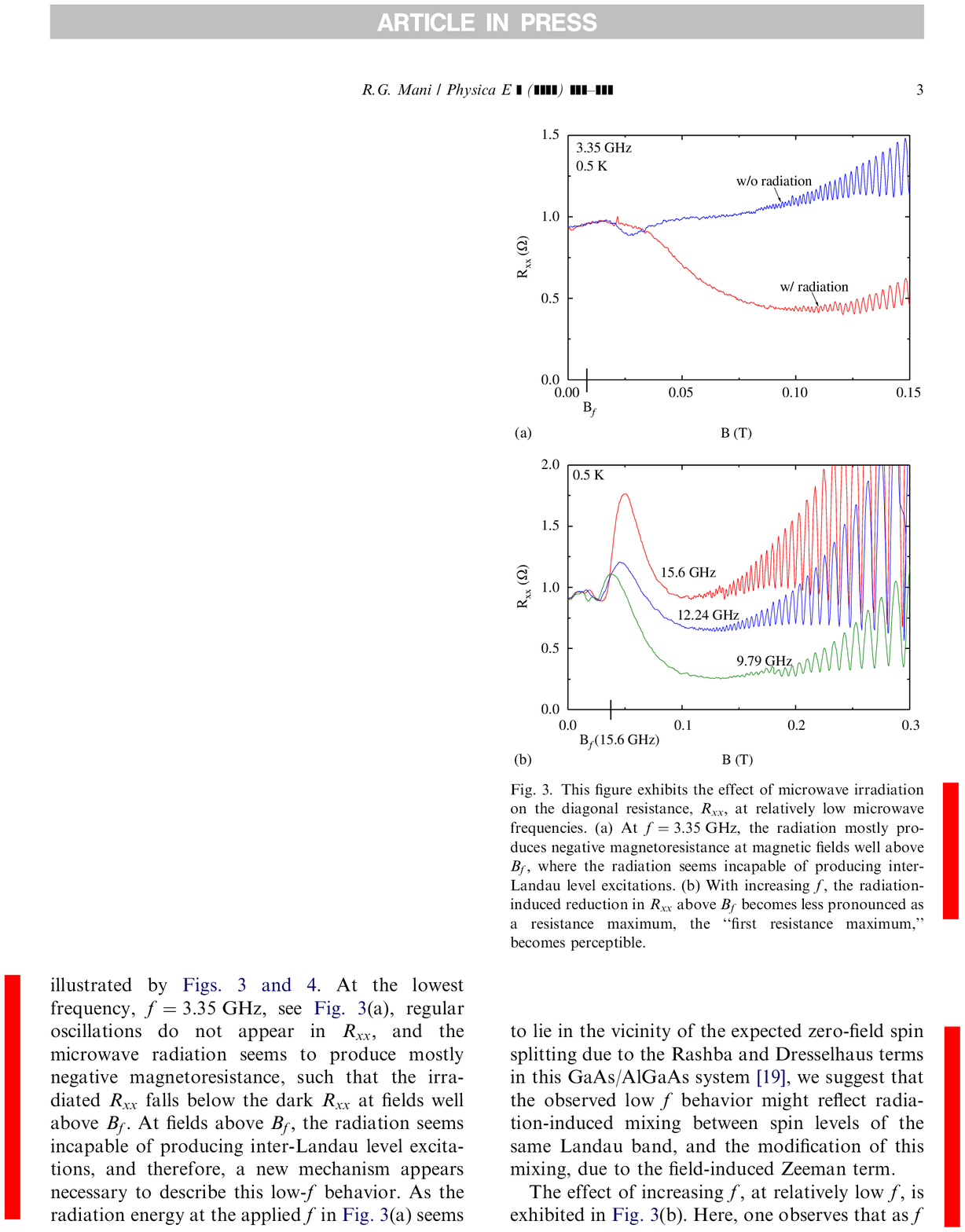}
 \label{1}
\end{center}
\end{figure}
\begin{figure}
\begin{center}
\includegraphics*[scale = 0.25,angle=0,keepaspectratio=true,width=4in]{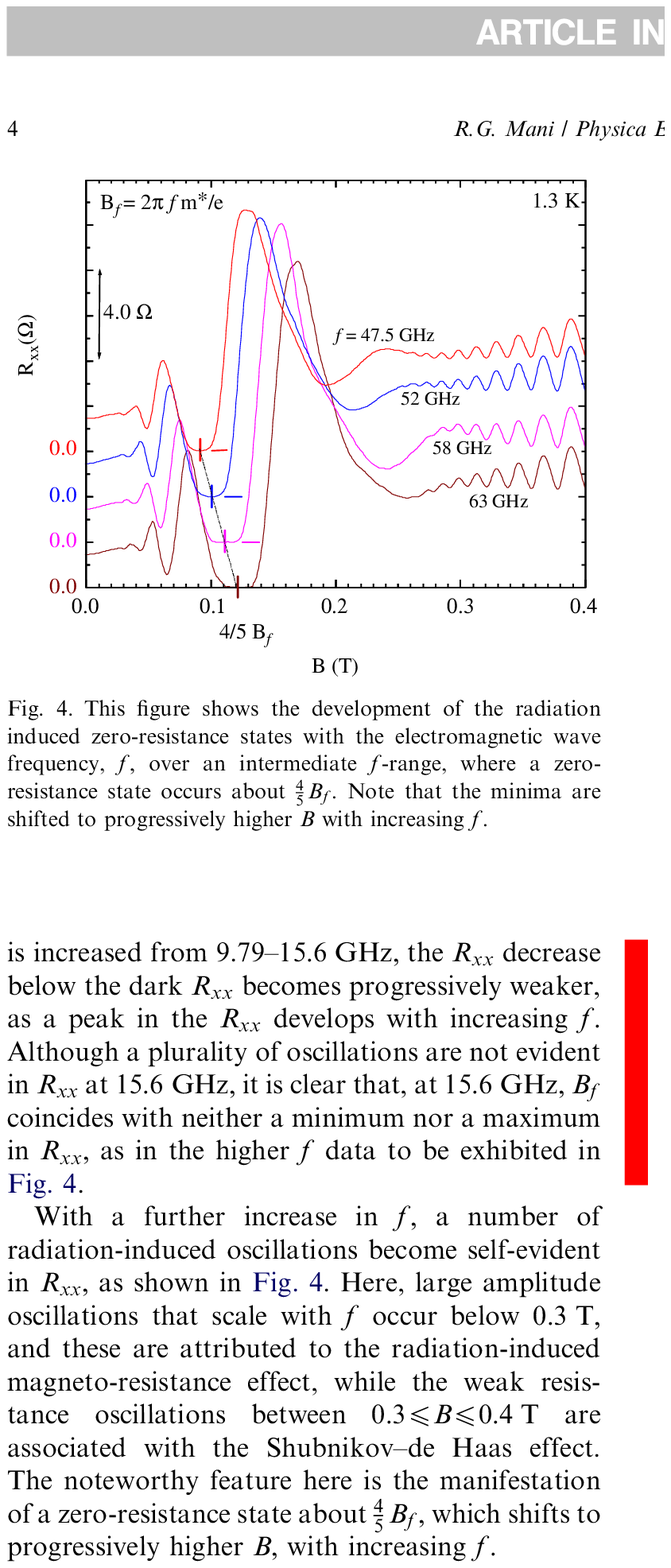}
 \label{1}
\end{center}
\end{figure}

In cond-mat/0409228,\cite{1} the claim to novelty seems to be that
the work represents the first conscious identification of a role
for \textit{intra Landau level transitions} in the microwave
induced resistance response in the $B > B_{f}$ limit.

An identification of a role for intra Landau level transitions in
the microwave induced magnetoresistance in the $B > B_{f}$ limit
has been made previously, however, at the 13th International
Winterschool on New Developments in Solid State Physics - Low
Dimensional Systems, in Mauterndorf (Salzburg), Austria.\cite{3}
An extract from the associated paper, where the microwave induced
magnetoresistance in the $B > B_{f}$ limit has been attributed to
spin-flip intra Landau level excitations, is provided in the
following two pages.\cite{3}

Perhaps, spin-flip intra Landau level excitations might constitute
a viable mechanism for understanding the observed photo-induced
magnetoresistance effect in the $B > B_{f}$ limit.\cite{3}
Certainly, spin-flip intra Landau level excitations remain
energetically plausible in the $B$ $>$ $B_{f}$ limit, the spin
splitting includes a $B$ variation due to the field induced Zeeman
term, and, the magnitude of the spin-splitting, upon including
spin-orbit effects, lies in the frequency (energy) range,\cite{4}
where the microwave radiation produces a resistance
reduction.\cite{3,5}

 \vspace{0cm}


\begin{thebibliography}{24}
\bibitem{5} The argument has been made that spin-effects need
not be considered here because "spin-splitting is not
resolved."\cite{1} Yet, it is not obvious that spin-splitting
needs to be "resolved" for spin to have a perceptible effect on
the radiation-induced magnetoresistance. Certainly, experiment has
already shown a radiation-induced magnetoresistance effect (for $B
< B_{f}$) even where Landau level splitting is apparently
"unresolved" by experiment, i.e., where SdH oscillations are not
observable, see, for example, Fig. 3(a) in R. G. Mani et al.,
Phys. Rev. Lett. 92, 146801 (2004).
\end{thebibliography}
\end{document}